# Intrinsic conductance of ferroelectric domain walls

Feng Yang

School of Materials Science and Engineering, University of Jinan, Jinan 250022, China

mse_yangf@ujn.edu.cn

## Abstract:

Ferroelectric domain walls hold great promise for innovative applications in ferroelectric devices. However, the underlying mechanisms behind the observed giant conductance of charged domain walls remain poorly understood. Using a first-principles approach that incorporates Boltzmann transport theory and the relaxation time approximation, we determine the carrier concentration, mobility, and conductivity of domain walls with head-to-head and tail-to-tail polarization orientations. Our systematic exploration reveals that the accumulation of carriers, particularly their concentration, plays a dominant role in the domain wall conductance mechanism. However, the observed conductance differences between head-to-head and tail-to-tail domain walls are primarily due to differences in carrier mobility. The width of the domain wall is a key factor determining the device scale. Our calculated domain wall width is significantly smaller than previously reported values. This method, not limited to a certain ferroelectric material, can be used for the optimization and application development of various domain wall materials and devices.

## I. Introduction

Ferroelectric materials are substances that exhibit a finite electric polarization even in the absence of an external electric field. They are characterized by having at least two stable states of polarization, which can be reversibly switched from one state to another through the application of an electric field.[1] Within a ferroelectric sample, there exist ferroelectric domains, which are regions with different orientations of the polarization vector. These domains can coexist and are separated by domain walls (DWs), as depicted in Figure 1.

Ferroelectric DWs represent discontinuous regions of polarized charges between domains and are often considered two-dimensional homointerfaces[2],[3] with a thickness of a few nanometers. The electric dipoles in ferroelectrics can be influenced and reoriented by external electric fields, allowing for the creation, erasure, or relocation of DWs within the material[4],[5]. Due to the emergence of distinct physical properties in spatially confined nano-regions, DWs have garnered attention for future applications in electronic, spintronic, and optoelectronic devices[6]. Several novel device concepts based on ferroelectric DWs have also been proposed, including ferroelectric DW pn junctions[7],[8], ferroelectric DW memory[9], ferroelectric DW Diodes[10], ferroelectric DW logic gates[9],[10] and ferroelectric DW memristors, etc.

In particular, the discovery of enhanced conductivity in the localized metal-insulator transition,[11] the conductivity at phase boundaries[12] in complex oxides, and the unexpected finding of the ferroelectric DW conductivity (in wide-bandgap ferroelectric semiconductors such as $BiFeO_3$,[13] $BaTiO_3$,[14] $LiNbO_3$,[15] and $PbZr_xTi_{1-x}O_3$[16]) have opened up new avenues for developing DW-based nanoelectronics and drawn extensive research interest. Notably, this enhanced electric conduction is not limited to thin films; it equally applies to millimeter-thick wide bandgap single crystals.[15],[17] Charged domain walls (CDWs)[18] exhibit uncompensated positive or negative polarization charges, creating a local electrostatic potential near the DWs. Typically, this potential is compensated by the redistribution of free carriers, further enhancing DW conductivity.[19] Moreover, at head-to-head (HH) CDWs, a large number of carriers accumulate, resulting in conductivity increases by several orders of magnitude. For instance, CDWs in the prototypical ferroelectric $BaTiO_3$ exhibit conductivity enhancements up to $10^9$ times that of the parent matrix.[14],[20] In $LiNbO_3$, this change leads to resistances more than 10 orders of magnitude lower compared to the surrounding bulk, with charge carriers confined to nanometer-wide sheets.[21] In such cases, conductivity correlates with the magnitude of polarization divergence at the wall, which can often be adjusted by changing the wall orientation. Carrier type can also be tuned: DWs that support HH polar discontinuities accumulate negative screening charges and exhibit n-type transport behavior, whereas tail-to-tail (TT) DWs accumulate positive screening charge and are found to be p-type[8],[7].

The physics responsible for generating DW conduction is still under debate and the reality is that details vary from one system to another.[22] Three main mechanisms have been identified, which can be classified as intrinsic or extrinsic effects.[22] Initially, it was ascribed to structural changes at the DW that produced a polarization discontinuity, leading to steps in the electrostatic potential, and a concomitant lowering of the band gaps.[2],[23],[24] Polar discontinuities at so-called "charged walls" create intense local fields, and the associated enhanced DW conduction appears to be more intrinsic in nature. Band bending is thought to be sufficient to either push the conduction band below the Fermi level (resulting in electron conduction within the wall) or raise the valence band above it (leading to hole conduction), depending on the sense of the polarization discontinuity (either "HH" or "TT").[25],[26],[27] Intrinsic effects can include a reduction in the electronic bandgap due to a subtly different electronic structure at DWs or a shift of the valence and conduction bands caused by DW-bound charges. DW conduction enhanced by such intrinsic effects is desirable whenever resistivity is controlled by the injection and deletion of DWs, leading to fast response times and avoiding degradation (for example, from pinning of DWs by point defects). In contrast, extrinsic DW conduction relies on point defects and is driven by the segregation of charged point defects toward or away from the DWs[16],[28],[29],[30], along with concomitant charge-compensating electrons or holes. It is important to note, however, that in real materials, all these mechanisms may be at play, and their deconvolution can be challenging. Furthermore, comprehensive information regarding the electronic bulk properties, including carrier densities (electron concentration n and hole concentration p) and mobilities (μ), is frequently unavailable, which complicates the assessment of domain-wall-specific responses.[31]

The reliable determination of the fundamental transport parameters of CDWs, particularly the charge carrier concentration and the mobility of the majority carriers, is of utmost interest.[21],[31] However, practically, the assessment of quantitative transport data related to DWs, such as the charge carrier type, concentration, and mobility, has been limited to only a few exemplary cases. Up to now, three measurement methods have been mainly reported experimentally: The first method is based on the scanning-probe-based approaches to analyze the Hall effect in DWs in improper ferroelectrics like $YbMnO_3$[32], $ErMnO_3$[33]. and proper ferroelectrics like $LiNbO_3$[8]. This method is mainly valid for extracting near-surface charge carrier densities and mobilities, but it cannot extract those inside the film. The second method involves extracting the $LiNbO_3$ carrier mobility from magnetoresistance analysis. However, this method requires limiting the shape of the DW to be conical in morphology.[34] The third method utilizes the classical van-der-Pauw four-point electrode configuration, analyzing "macroscopic" Hall effect measurements of materials to quantify two-dimensional carrier densities and Hall mobilities.[35],[21] Although this method has improved, it still has an error of more than 10%.[35],[21] It is evident that experimental work in this area is still insufficient. In fact, even determining the intrinsic conductance of ferroelectric DWs remains challenging through experiments. In ferroelectric thin film devices, there is a Schottky barrier at the thin film−electrode interface, making it difficult to judge whether DW conduction or the interfacial Schottky barrier dominates the conductive behavior.[36] This challenge is particularly pronounced in thicker films. Additionally, the DW current varies from device to device within the same type of ferroelectric layer due to the interference of randomly distributed defects, resulting in poor reproducibility and an inconsistent on/off current ratio.[36] The defect concentration on the surface of the thin film can not only alter the interfacial barrier state but also significantly affects the defect-mediated charge injection and trapping behavior under an external electric field, deeply influencing the DW current.[36] Therefore, theoretical research on the charge carrier concentration and the mobility of the majority carriers is highly anticipated.

Here, we seek to understand the transport mechanism of ferroelectric DWs. Predicting the ultimate magnitude of conductivity and understanding its transport mechanism are critical to realizing the optimal performance of these ferroelectric DW devices. In this study, a generalized Boltzmann transport method based on first principles is employed to predict the charge carrier mobility and conductivity of the ferroelectric DWs associated with domains having HH and TT polarization orientations. In our simulations, all the physical quantities are intrinsic limits of the system without defects or dopant atoms. We calculated the concentration and mobility of the carriers, thus determining the conductivity of the ferroelectric DWs. The transport mechanism has been systematically investigated using first-principles calculations for the first time. These results provide important insights into the understanding of charge carrier transport in ferroelectric DW devices.

## II. METHODS

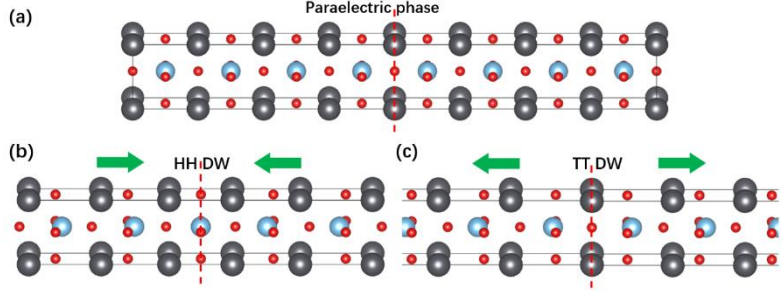

Figure 1. (Color online) schematic representation of (a) paraelectric phase, (b) head-to-head domain wall, and (c) tail-to-tail domain wall. Atoms are represented by sphheres: O in red, Pb in black, and Ti in blue. The green arrows indicate the direction of polarization.

First-principle calculations are performed within density functional theory (DFT)[37],[38] using the Vienna ab initio simulation package.[39],[40],[41] Exchange and correlation effects are treated within the local density approximation (LDA)[42] to simulate the 8-unit periodic DW structure containing both HH and TT 180° DWs in PbTiO$_3$. Additionally, the generalized gradient approximations (GGA)[43] with the Perdew-Burke-Ernzerhof (PBE) parameterization are also employed as the exchange-correlation potential to calculate the polarization in the ferroelectric domain and the electronic transport properties of the DWs. The valence electron configurations of O ($2s^2\ 2p^4$), Ti ($3s^2\ 3p^6\ 3d^2\ 4s^2$), and Pb ($5s^2\ 5p^6\ 5d^{10}\ 6s^2\ 6p^2$) have been taken into consideration. The cutoff energy of the plane-wave is set at 630 eV.

To model both domain structures, we employed a tetragonal (1×1×n) supercell, periodically repeated in space, of the composition (PbO-TiO$_2$)$_n$ with n = 8, as depicted in Figure 1(a). Initially, we defined an ideal structure by stacking n unit cells of PbTiO$_3$ along the [001] direction in a tetragonal ferroelectric configuration, with polarization directions on both sides converging toward the midpoint. By constraining other parts of the structure and subsequently relaxing the DWs, we obtained a stable crystal structure, as illustrated in Figure 2(a). For the end faces of the structure, we applied periodic boundary conditions.

The electronic transport properties are implemented as follows, utilizing Boltzmann transport theory and relaxation time approximation[44]. In a semiconductor system, the conductivity $\sigma$ is given by $\sigma = qn\mu_n + qp\mu_p$, where $\mu_n(\mu_p)$ represents the mobilities of electrons (holes) and n(p) denotes the electron (hole) concentration.

To calculate the carrier concentration, we choose the Fermi-Dirac distribution for the occupation of the one-particle Kohn-Sham electronic eigenstates at room temperature. The concentrations n (p) of electrons (holes) can be calculated using the following formulas:[45]

$$n(T,u) = \frac{2}{V} \iint_{BZ} f_0(T,\varepsilon,u) D(\varepsilon) d\varepsilon \tag{1}$$

$$p(T,u) = \frac{2}{V} \iint_{BZ} [1 - f_0(T,\varepsilon,u)] D(\varepsilon) d\varepsilon \tag{2}$$

where $V$ is the unit cell volume, $f_0$ is the Fermi-Dirac distribution, $u$ is the chemical potential, $\varepsilon$ is the electron energy, $D(\varepsilon)$ is the density of states (DOS).

The carrier mobility quantifies how quickly an electron or hole can travel in a metal or semiconductor under the influence of an external electric field E. Within the framework of the effective mass theory, the Drude model describes electron mobility as $\mu = e\tau/m^*$, where $m^*$ is the electron's effective mass and τ is the scattering time.[46] This relationship highlights the importance of accurately calculating both the scattering rate (1/τ) and the carrier effective mass ($m^*$) for predictive theories of carrier mobility. Here, the mobility in the $\beta$ direction can be written as $\mu_\beta = e\langle\tau_\beta\rangle/m^*$, where $\langle\tau_\beta\rangle$ is the average relaxation time.[47] Carrier mobility is calculated using the Boltzmann transport theory, the relaxation time approximation, and the effective mass theory. Deformation potential (DP) theory can be applied to determine the relaxation time.[48] The periodic structure of the five-atom unit cell of $PbTiO_3$ is used to calculate the effective mass and mobility of carriers. Structural distortions and residual strains of the unit cell are extracted from the regions of HH or TT DWs. The energy convergence of these atomic structures has been verified.

The scattering rate of an electron from an initial state $(i,\vec{k})$, where $i, j$ are the band indexes and $\vec{k}$, $\vec{k}'$ are the electron wavevectors, to a final state $(j,\vec{k}')$ is described by Fermi's golden rule as[47]

$$\frac{1}{\tau(i,\vec{k})} = \frac{2\pi}{\hbar} \sum_{\vec{k}',j} |M(i\vec{k},j\vec{k}')|^2 \delta[\varepsilon_i(\vec{k}) - \varepsilon_j(\vec{k}')] \tag{3}$$

Where τ represents the relaxation time in which electrons lose their momentum due to scattering from material defects, impurities, or lattice vibrations.[46] $\hbar$ is the reduced Planck's constant, ε is the electron energy, δ is the Dirac delta function and $M$ is the scattering matrix element. The matrix element $M(i\vec{k},j\vec{k}') = \langle j,\vec{k}'|\Delta V|i,\vec{k}\rangle$ describes scattering from state $(i,\vec{k})$ to state $(j,\vec{k}')$ by the deviation potential arising from atomic displacement associated with phonons or the perturbation potential caused by defects or impurities. This equation applies to perfectly elastic scattering[44], where electrons maintain their energy during the process. When considering relaxation time in the external field direction $\beta$ (i.e., $\tau = \tau_\beta$), the right-hand side of the above equation must be multiplied by $[1 - v_\beta(j,\vec{k}')/v_\beta(i,\vec{k})]$, which describes the scattering angle weighting factor[49]. Here, $v_\beta = \partial\varepsilon(\vec{k})/\hbar\partial\vec{k}_\beta$ represents group velocity. For a spherical energy surface $\varepsilon(\vec{k}) = \hbar^2 k^2/(2m^*)$, the weighting factor becomes (1- cos$\theta$), where $\theta$ is the angle between wave vectors. The main challenge lies in calculating the scattering matrix element. Here, we focus on the dominant scattering of a thermal electron or hole by an acoustic phonon within the DP theory.

The deformation potential theory, proposed by Bardeen and Shockley[48] in the 1950s, describes charge transport in non-polar semiconductors. The deformation potential theory describes the electron-acoustic phonon interactions, assuming that the local deformation produced by the lattice waves is close to that in homogeneously deformed crystals. Because the electron velocity with energy $k_B$T at 300 K is about $10^7$ cm s$^{-1}$, and the corresponding wavelength ($\lambda = h/(mv)$) is 7 nm, which is much larger than the lattice constant, the electron is primarily scattered by the acoustic phonons. The DP theory assumes that the lattice potential perturbation due to thermal motions has a linear dependence on the relative volume change. When considering only acoustic phonons, the matrix element for the scattering of an electron (or hole) from Bloch state $|i,\vec{k}\rangle$ to $|i,\vec{k}'\rangle$ is expressed as:

$\left|M(i\vec{k},j\vec{k}')\right|^2 = \left|j,\vec{k}'|\Delta V|i,\vec{k}\right|^2 = \left(E_1^i\right)^2 q^2 a_{\vec{q}}^2/N$, where $\vec{q} = \pm(\vec{k}' - \vec{k})$. At high temperature, when the lattice waves are fully excited, the amplitude of the wave is given by $a_{\vec{q}}^2 = k_B T/(2mq^2 v_a^2)$ according to the uniform energy partition theory. Here, $\vec{v}_a$ is the velocity of the acoustic wave, and m is the total mass of the lattice per unit volume. Finally, the average scattering probability becomes[47],[48], $\langle|M(i\vec{k},j\vec{k}')|^2\rangle = k_B T (E_\beta^i)^2/C_\beta$, where $C_\beta = \rho v_a^2 = Nmv_a^2$ is the elastic constant for longitudinal strain in the propagation direction of the LA wave ($\beta$). $E_\beta^i$ is the DP constant of the $i$-th band. Here, we assume that the scattering matrix element is independent of state $\vec{k}$ or $\vec{k}'$, and that the charge transport direction (electric field direction) is parallel to the wave vector of the LA phonon.

The DP constant $E_\beta^i$ and the elastic constant $C_\beta$ are determined through electronic structure calculations.[50],[51],[52],[47] The elastic constant $C_\beta$ was calculated using the total-energy versus strain approach, where the crystal volumes were isotropically changed by ±1 %. The deformation potential $E_\beta^i$ represents the shift of the band edge (either the valence band maximum or the conduction band minimum) per unit strain.

The DOS and the partial density of states (PDOS) were used to calculate the carrier concentration, while band structure calculations yielded the effective mass. To calculate mobility, you must first determine the effective mass. Under an external electric field, which is treated classically, the effective mass of a charge carrier is defined as $m^*_{\alpha\beta} = \hbar^2[\partial^2 E_n(\vec{k})/\partial k_\alpha \partial k_\beta]^{-1}$, where indices α and β denote reciprocal components, and $E_n(\vec{k})$ is the dispersion relation for the n-th band. Derivatives can be evaluated numerically using a finite difference method, such as a five-point stencil,[53] with an error of the order of O($h^4$). Unlike the commonly used single parabolic band approximation,[47] our method is not limited to the spherical energy surface.

## III. Results and discussion

After relaxing the DW structure of eight unit cells, we observed that the most stable phase is the centrosymmetric unpolarized structure.[54] Fortunately, by partially constraining the domain structure, we obtained a meta-stable 8-unit periodic DW structure containing HH and TT 180° DWs in PbTiO$_3$, as shown in Fig2(a). The structural parameters of the meta-stable DW structure obtained after relaxation are a= 3.822 Å, b= 3.822 Å, and c= 33.778 Å. The numbers in Fig2(a) correspond to the ordering of each layer in the crystal structure below. For example, 1 corresponds to the PbO layer on the left, and 2 corresponds to the TiO$_2$ layer.

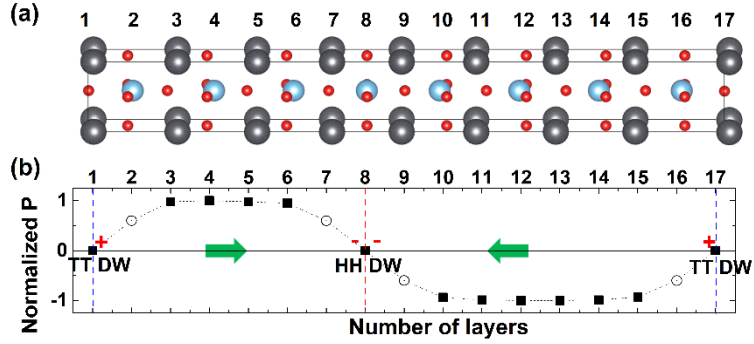

Figure 2. (Color online) (a) Relaxed 8-unit periodic DW structure containing HH and TT 180° DWs in PbTiO3; (b) Local layer-by-layer normalized effective polarization profiles for the periodic DW structure in (a). Black squares represent normalized polarization of each layer, computed from Born effective charges. Open circles are not calculated but are estimates. The vertical dashed line marks the position of the DW (red: HH DWs; blue: TT DWs). Vertical dotted red lines denote the positions of the TiO$_2$ layers, while the vertical dashed blue line marks the position of the PbO layers. The flat value of the polarization at the center of each domain amounts to P = ±136.53 μC/cm$^2$. Green arrows represent the direction of the macroscopic polarization within each domain. The + and − correspond to the sign of the polarization-induced charge concentration at TT and HH DW.

In Figure 2(b) we plot the layer-by-layer effective polarization profile in the 8-unit periodic DW structure containing HH and TT 180° DWs in PbTiO$_3$. The layer polarizations have been calculated using an approximate expression based on the effective Born charge method. The polarization profile is remarkably flat within the interior of the domains, characterized by a constant value of 136.53 μC cm$^{-2}$. While this value is larger than previously reported (ranging from 45μC cm$^{-2}$ to 92 μC cm$^{-2}$)[54],[55], it is still far smaller than that of super-tetragonal PbTiO$_3$ thin films, which exhibit giant polarization as large as 236.3 μC cm$^{-2}$.[56] At the DWs, we observe a spatial variation of *P*, resulting from localized strong non-homogeneous atomic distortions. The extension of these regions with nonuniform polarizations, spanning approximately two cells, allow us to quantify the width of the DWs. In our relaxed metastable structure, a DW width with a scale of only 2 unit cell scale is obtained, which is much smaller than previous experimental and theoretical results (indicating a critical thickness of 16 unit cells).[54] Specifically, HHDW corresponds to layers 6, 7, 8, 9, and 10, while TTDW corresponds to layers 1, 2, 3, 15, and 16. We predict that the thickness of the DW is approximately two unit cells, which aligns with the same order of magnitude as the previously predicted value of $10^{-7}$ cm[57]. Additionally, it is important to note that the smaller the band gap of the material, the shorter the critical thickness required by the electrostatic potential to close the gap, leading to the earlier appearance of CDWs.[54] Worth mentioning is that the LDA method we used generally underestimates the band gap of the material, which may consequently underestimate the critical thickness of DWs.

These 180° DWs, where the wall is perpendicular to the polarization, give rise to polarization-induced bound charges, that lead to the local concentration of a large electrostatic energy that destabilizes this configuration. The stabilization of the domains observed in Figure 2 implies that the positive (for the HH) and negative (for the TT) polarization charges at the DW have been screened. Since our calculations do not consider the presence of defects or dopants, the ferroelectric polarization near the $PbTiO_3$ DW can be screened by electronic reconstruction and the formation of two-dimensional metallic gases at the interfaces[54]. By construction, the central $TiO_2$ layer serves as a mirror symmetry plane. This differs from the previous analysis[54], where the central layer they calculated was a PbO layer, and their calculations were based on an independent DW structure containing a vacuum end face[54]. Our calculations demonstrate that for HHDW, the symmetry center is the $TiO_2$ layer; for TTDW, the symmetry center is the PbO layer. The metallization of the DW is due to neutralization by free carriers, resulting in the formation of thin quasi-two-dimensional metallic layers. This has been confirmed in the subsequent analysis of the carrier concentration at the domain walls.

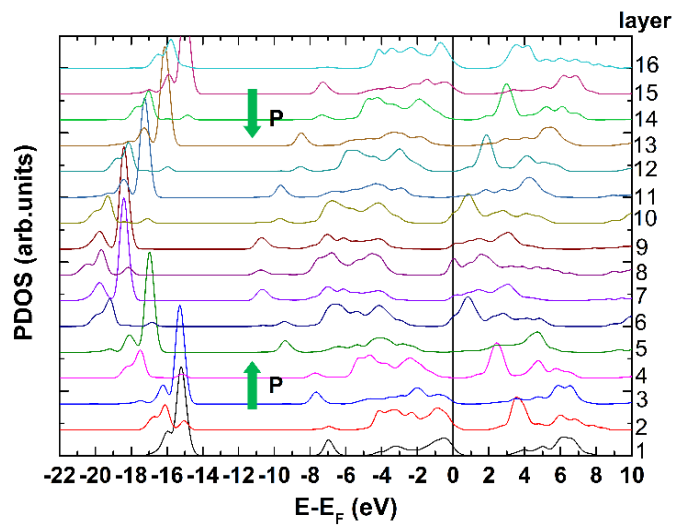

Figure 3. (Color online) Layer-by-layer PDOS (solid black curves with gray shading) for the 8-unit periodic

DW containing HH and TT 180° DWs in PbTiO$_3$. The bottom curve corresponds to the first layer, and the top one corresponds to the 16th layer. The Fermi level is located at zero energy (vertical black line). The polarization direction is indicated by the green arrow.

Figure 3 shows the layer-resolved PDOS on the atomic orbitals of Ti and O in the TiO$_2$ layers and the atomic orbitals of Pb and O in the PbO layers of the different PbTiO$_3$ unit cells. The energy scale is in eV, with the Fermi level ($E_F$) set to zero. As indicated on the right side of Figure 3, the bottom curve corresponds to the first layer, and the top one to the 16th layer. A well-defined energy gap is observed in all layers, although the Fermi level position varies.

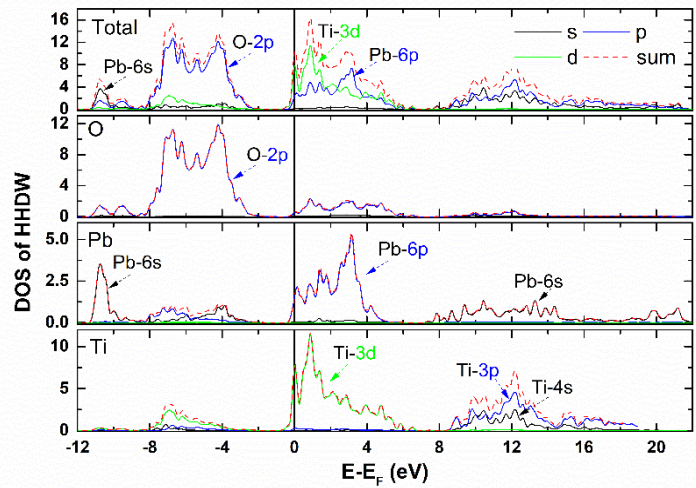

Figure 4. The DOS for head-to-head domain walls in PbTiO$_3$, along with the PDOS for each type of atom within the domain wall.

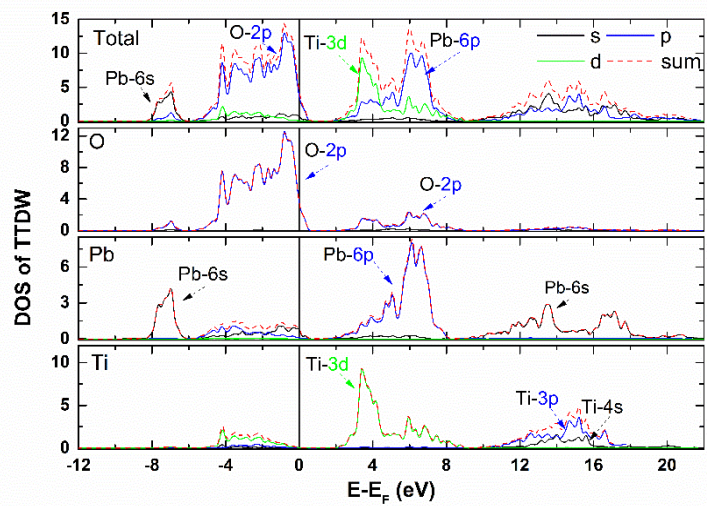

Figure 5. The DOS for tail-to-tail domain walls in PbTiO$_3$, along with the PDOS for each type of atom within the domain wall.

Figure 4 shows the DOS at the head-to-head DWs in PbTiO$_3$, along with the PDOS for each type of atom within the domain wall. In the region from -12 eV to -2 eV, below the Fermi level, these bands mainly originate from O-2p orbitals. Above the Fermi level, these bands result from the hybridization of O-2p, Ti- 3d and Pb-6p orbitals. Notably, at the HH DW, the Fermi level lies above the bottom of the conduction band, leading to system degeneracy. Figure 5 depicts the DOS at the tail-to-tail DWs in $PbTiO_3$, along with the PDOS for each type of atom within the domain wall. In the energy range from -8 eV to zero, below the Fermi level, these bands primarily arise from O-2p orbitals. Above the Fermi level, these bands result from the hybridization of O-2p, Ti-3d and Pb-6p orbitals. Interestingly, at the TT DW, the Fermi level enters below the valence band top, also contributing to system degeneracy.

The local metallization of the DWs becomes evident when considering the position of the valence-band top and the conduction-band bottom within these layers relative to the Fermi level of the entire structure (taken as zero in Figure 4 and Figure5). In the HH domain configuration (Figure 4), the non-vanishing PDOS at the Fermi level indicates the electronic population of the conduction-band bottom in the $PbTiO_3$ unit cells at the HH DW. This free electronic charge has been transferred from the TT DW, as evidenced by the generation of holes at the TT DW (observed through the crossing of the Fermi level with the PDOS at the valence-band top for the TT DW). Similarly, in the TT domain configuration (Figure 5), we observe a similar behavior, but with the roles of electrons and holes interchanged. Specifically, holes are generated at the top of the valence band in the TT DW, while electrons populate the bottom of the conduction band in the HH DW.

The PDOS of the conduction and valence bands converges fairly quickly to the bulk curve when moving away from the interface, and the PDOS vanish at the Fermi level, indicating local insulating behavior. Furthermore, the rigid shift observed in the PDOS of each layer suggests the presence of an internal remnant depolarization field. This field electrostatically confines the free charges to the metallic region within the DWs[58] and plays a crucial role in determining the atomic orbitals involved in the electronic reconstruction[54], as evidenced by the bending of the energy bands depicted in Figure 6.

As illustrated in Figures 3, 4, and 5, carrier behavior depends on the DOS near the band edges according to semiconductor energy band theory and the Fermi-Dirac distribution. Specifically, the concentration of electrons is determined by the DOS near the conduction band minimum, while the concentration of holes depends on the DOS near the valence band maximum. Further insights into the origin of these electron and hole charges can be obtained by decomposing their orbital contributions, as shown in Figures 4 and 5. Our findings from Figure 4 reveal that the HH electron gas originates primarily from Ti-3d orbitals, with a minor contribution from Pb-6p orbitals. Conversely, the TT hole gas has a primary contribution from the O-2p orbitals in both the PbO and $TiO_2$ planes. Our calculations reveal that electron reconstruction occurs due to the screening of polarization-induced charges at the 180° DWs.[31],[6] The polarization discontinuity at the DW serves as the driving force for electronic reconstruction. Additionally, charge transfer between the HH DW and the TT DW acts as an effective screening mechanism, stabilizing the two-DWs configurations. Both HHDW and TTDW exhibit localization effects due to the remnant electric potential energy within the domains. These potentials, shaped like a ∨ for HH and ∧ for TT (as shown in Figure 6), confine the respective carriers. Within the energy potential well, a stable, degenerate electron (hole) gas forms to screen polarization divergences in HH (TT) DWs[59]. This scenario is characterized by the presence of a substantial amount of free charge populating the band edges.

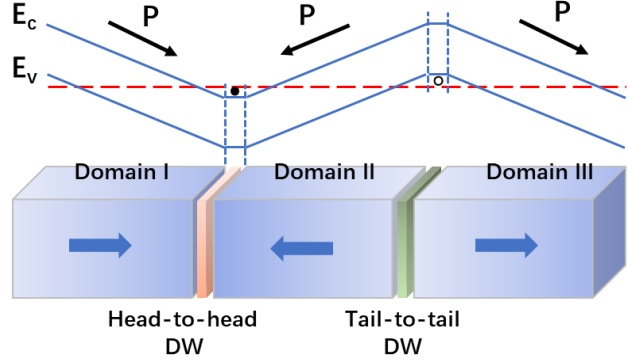

Figure 6. Energy band diagram of the periodic DW containing head-to-head and tail-to-tail 180° DWs in PbTiO$_3$. Ec denotes the bottom of the conduction band, and Ev denotes the top of the valence band. P represents the polarization, with the arrow indicating its direction. Black dots represent accumulated free electrons, and open circles represent accumulated holes.

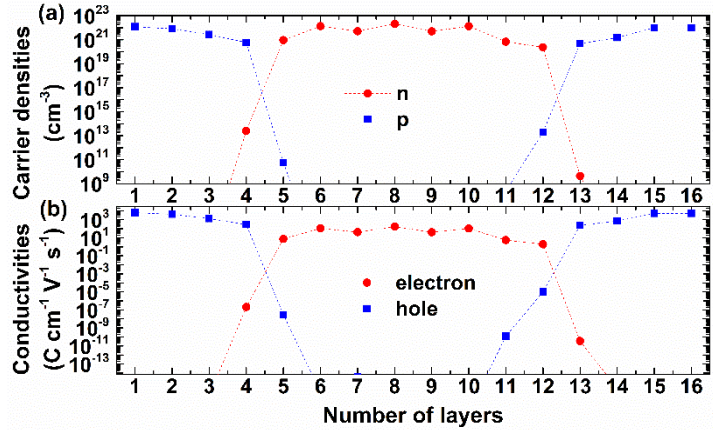

Figure 7. (Color online) Calculated free carrier densities (a) and conductivities (b) for each layer in the 8-unit periodic DW containing HH and TT 180° DWs in PbTiO$_3$. In (a), the blue squares represent the hole concentration (p) and the red dots represent the free electron concentration (n) in each layer. The x-axis indicates the atomic layer number within the periodic domain structure. In (b), the blue squares represent the hole conductivity and the red dots represent the free electron conductivity in each layer. The green arrows represent the polarization configuration.

To estimate the amount of charge transferred from the TT DW to the HH DW for the purpose of screening depolarizing fields, we further calculated the carrier (electron and hole) concentration in each layer of PbO$_2$ and TiO$_2$, as depicted in Figure 7(a). For electronic concentration calculations, we employed the Fermi-Dirac distribution to describe the occupation of one-particle Kohn-Sham electronic eigenstates, assuming a temperature of 300 K for both HH and TT domain structures. Our calculation results indicate that a carrier concentration of $10^{21} \sim 10^{22}$ cm$^{-3}$ electrons (holes) per cubic centimeter is necessary to effectively screen HH (TT) walls in PbTiO₃ (a proper ferroelectric). Notably, this value aligns with previously reported carrier concentrations, which were also in the order of magnitude of $10^{21}$ cm$^{-3}$ [57].

Confirming the non-local charge neutrality observed at the HH and TT DWs, we can integrate the profiles of the average free charge densities along the (001) direction. This integration provides the free carrier concentration per unit volume: n (electrons) and p (holes). The HH DW exhibits an additional electron concentration that populates the bottom of the conduction band. This concentration is represented as $n^{\mathrm{HH}}$ = 1.1632×10$^{22}$ electrons/cm$^3$ [Figure 7(a)]. These electrons are transferred from the TT DW, where the integration of the free hole charge in Figure 7(a) yields a value of $p^{\mathrm{TT}}$ = 8.6941×10$^{21}$ holes/cm$^3$. The global charge neutrality condition requires that $n^{\mathrm{HH}} V_{\mathrm{DW}}^{\mathrm{HH}} = p^{\mathrm{TT}} V_{\mathrm{DW}}^{\mathrm{TT}}$, where $V_{\mathrm{DW}}$ corresponds to the volume of the DW. The extra holes at the TT DW equals the number of additional electrons populating the bottom of the conduction bands at the HH DW[54].

The electronic transport properties are calculated using Boltzmann transport theory, effective mass theory and relaxation time approximation. Our calculated electron mobility ($\mu_n$) of HH DW electrons is 5.0877 × 10$^{-2}$ cm$^2$/V·s, which is consistent in order of magnitude with the free carrier mobility of 1.49 × 10$^{-2}$ cm$^2$/V·s reported by Chen in the BiFeO₃ CDW region[36]. The mobility ($\mu_p$) of TT DW holes we calculated is 3.0902 cm$^2$/V·s, which is much larger than the mobility of electrons at HH DW. In Figure 7(b), we present the calculated conductivities for each layer of the 8-unit periodic DW containing both HH and TT 180° DWs in PbTiO₃. Specifically, for PbTiO₃ CDWs, the average electron conductivity at HH DW is 9.4813×10$^3$ S/m, while the average hole conductivity at TT is 5.0454×10$^5$ S/m. The calculated conductivity of the DWs reaches 10$^3$~10$^5$ S/m, significantly higher than the conductivity of typical semiconductor Si (4.3478×10$^{-4}$ S/m). This result further confirms good DW conduction, although it remains smaller than the conductivity of typical metal Au (4.11×10$^7$ S/m). As depicted in Figure 7(b), the hole conductance from TT DWs is ten times that of the electron conductance from HH DWs. This difference arises from the larger hole mobility observed at the TT DW. Additionally, it has been reported that the TT DW conductance in ErMnO₃ DWs significantly exceeds the HH conductance[25].

A comprehensive analysis reveals that domain wall conductance originates from the accumulation of charge carriers, with carrier concentration playing a predominant role in the conduction mechanism. However, the conductance difference between HH DWs and TT DWs primarily stems from carrier mobility.

In CDWs, polarization-induced bound charges lead to a significant and unfavorable electrostatic energy[60]. These DWs exhibit behavior reminiscent of metals, but they are either impractically unstable or must be pinned by defects.[14] However, naturally locked ferroelectric DWs - those with stable charged structures - have been observed only in improper ferroelectrics such as YMnO$_3$ (ref. [61]) and ErMnO$_3$ (ref. [25]). Researchers like Sifuna et al.[54] and Gureev et al.[59] propose that CDWs can form in defect-free ordinary ferroelectrics when the sample thickness is sufficiently large. In contrast, our calculations reveal that charged HH and TT DWs at the nanoscale can exist even without external defects. These stable CDWs, uncompensated by mobile defects, are referred to as intrinsic CDWs in ferroelectrics.[57],[59],[62] While individual HH and TT DWs become unstable during large-scale energy relaxation, periodic structures with alternating HH and TT DWs can exist in a metastable state. This metastable state remains stable as long as perturbations do not exceed a certain energy threshold. In ordinary ferroelectrics like PbTiO$_3$, the existence of such 'strongly' CDWs critically depends on near-perfect compensation of polarization charges. The charge compensation in stable CDWs of ordinary ferroelectrics results in the formation of a degenerate quasi-two-dimensional electron gas with a metallic-like carrier concentration. This high concentration could lead to significant steady-state metallic conductivity and other properties associated with electron gas confinement within the energy well formed by the polarization charges.[14]

The next challenge lies in developing CDW devices for practical applications. A technique called frustrated poling can be used to create stable CDWs during device fabrication.[63],[14] However, the choice of electrode materials is crucial as it needs to consider the contact potential difference between the ferroelectric and the metal electrodes (i.e., the work function difference between the metal and the ferroelectric)[57]. Additionally, the coupling between ferroelectric-ferroelastic domain states can also be harnessed for designing and constructing stable CDWs.[14]

## IV. Inclusion

Our DFT simulations demonstrate the stabilization of 180° HH and TT CDWs within a periodic DW configuration in an ideal PbTiO$_3$ system (free of dopants and vacancies). This stabilization is facilitated by electrostatic screening provided by free carriers. When periodic DWs exceed a critical size (in our case, 8-unit cells), electronic reconstruction occurs. This process involves the transfer of electron charges from TT DWs to HH DWs, resulting in the formation of quasi-two-dimensional electron and hole gases. Using a first-principles DFT approach that incorporates Boltzmann transport theory and the relaxation time approximation, we determined the carrier concentration, mobility and conductivity for both types of DWs. In the 8-unit cell PbTiO$_3$ system, the average electron conductivity at the HH DW is 9.4813×10$^3$ S/m, with an electron concentration (n) of 1.1632×10$^{22}$ electrons/cm$^3$ and mobility ($\mu_n$) of 5.0877 × 10$^{-2}$ cm$^2$/V·s. Conversely, the average hole conductivity at the TT DW is 5.0454×10$^5$ S/m, with a hole concentration (p) of 8.6941×10$^{21}$ holes/cm$^3$ and mobility ($\mu_p$) of 3.0902 cm$^2$/V·s. These values confirm good DW conduction. Furthermore, our systematic investigation reveals that the accumulation of carriers, particularly their concentration, plays a dominant role in the DW conductance mechanism. However, the observed difference in conductance between HH and TT DWs primarily arises from the disparity in carrier mobility. These findings provide valuable insights into the carrier transport mechanisms within ferroelectric DW devices. In summary, we propose a first-principles method for calculating the conductivity of ferroelectric DWs, applicable to various ferroelectric materials and their DW devices. This method offers valuable insights into the internal mechanisms governing these devices, facilitating their optimization and application development.

# Acknowledgments

Funding: The work was supported by funding from the Natural Science Foundation of Shandong Province (No. ZR201702120113),